\documentclass[iop,apj]{emulateapj}

\usepackage{graphicx}
\usepackage{color}
\usepackage{amsmath}
\usepackage{xspace}
\usepackage{hyperref}
\usepackage{xcolor}

\hypersetup{
colorlinks,
linkcolor={red!50!black},
citecolor={black},
urlcolor={blue!80!black},
pdfauthor= Elisa Quintana, 
pdftitle = Giant Impacts on Earth-like Worlds
}

\newcommand{\mearth}{M\ensuremath{_{\oplus}}\xspace}

\newcommand{\mimp}{\ensuremath{m_{imp}}\xspace}
\newcommand{\mtar}{\ensuremath{m_{tar}}\xspace}
\newcommand{\vimp}{\ensuremath{v_{imp}}\xspace}

\newcommand{\rearth}{\ensuremath{R_{\oplus}}\xspace}

\newcommand{\gcm}{\ensuremath{\rm{g\,cm^{-3}}}\xspace}

\newcommand{\rtar}{\ensuremath{r_{tar}}\xspace} 	
\newcommand{\rimp}{\ensuremath{r_{imp}}\xspace}

\slugcomment{Revision \today}

\shorttitle{The Frequency of Giant impacts on Earth-like Worlds}
\shortauthors{Quintana et al.}

\begin{document}
\title{The Frequency of Giant Impacts on Earth-like Worlds}

\author{Elisa V. Quintana\altaffilmark{1,2}, Thomas Barclay\altaffilmark{1,3}, William J. Borucki\altaffilmark{1}, Jason F. Rowe\altaffilmark{1,4} and John E. Chambers\altaffilmark{5}}

\altaffiltext{1}{NASA Ames Research Center, Moffett Field, CA 94035}
\altaffiltext{2}{NASA NPP Senior Fellow}
\altaffiltext{3}{Bay Area Environmental Research Institute, 625 2nd St. Ste 209, Petaluma, CA 94952}
\altaffiltext{4}{SETI Institute, 189 Bernardo Ave, Mountain View, CA 94043}
\altaffiltext{5}{Department of Terrestrial Magnetism, Carnegie Institution for Science, 5241 Broad Branch Road NW, Washington, DC 20015, United States}

\begin{abstract}
The late stages of terrestrial planet formation are dominated by giant impacts that collectively influence the growth, composition and habitability of any planets that form. Hitherto, numerical models designed to explore these late stage collisions have been limited by assuming that all collisions lead to perfect accretion, and many of these studies lack the large number of realizations needed to account for the chaotic nature of $N$-body systems. We improve on these limitations by performing 280 simulations of planet formation around a Sun-like star, half of which used an $N$-body algorithm that has recently been modified to include fragmentation and hit-and-run (bouncing) collisions. We find that when fragmentation is included, the final planets formed are comparable in terms of mass and number, however their collision histories differ significantly and the accretion time approximately doubles. We explored impacts onto Earth-like planets which we parameterized in terms of their specific impact energies. Only 15 of our 164 Earth-analogs experienced an impact that was energetic enough to strip an entire atmosphere. To strip about half of an atmosphere requires energies comparable to recent models of the Moon-forming giant impact. Almost all Earth-analogs received at least one impact that met this criteria during the 2 Gyr simulations and the median was three giant impacts. The median time of the $\emph{final}$ giant impact was 43 Myr after the start of the simulations, leading us to conclude that the time-frame of the Moon-forming impact is typical amongst planetary systems around Sun-like stars.
\end{abstract}

 
\section{Introduction}\label{sec:intro}
The final stages of terrestrial planet formation are characterized by a myriad of gravity-dominated collisions among bodies in a protoplanetary disk. These impacts can lead to a wide range of outcomes depending on the masses, impact speed, impact angle, composition and number of bodies involved.  Taken together, collisions can largely influence a planet's growth, stability, bulk composition and habitability. 

Gravitational $N$-body algorithms have been widely used for decades to numerically model the late stages of terrestrial planet formation.  Numerical integrations of these late stages - in which planets form via pairwise accretion from bodies in a protoplanetary disk - face several major complexities. First, $N$-body simulations that include interparticle self gravity are computationally intensive, and the computation time scales as the square of the number of bodies ($N^2$). This is one reason why nearly all previous $N$-body models have assumed perfect accretion (meaning two bodies that collide will stick together and conserve mass and momentum), limiting the maximum $N$ to the initial number of bodies in the protoplanetary disk. Collisional fragmentation, however, can increase $N$ so has been neglected in these models. Secondly, $N$-body systems with $N$ $>$ 2 are chaotic and thus a large number of simulations of a given system, with very small changes in the initial conditions, are required in order to gain meaningful statistical results. 

Despite these limitations, numerical $N$-body simulations that assumed perfect accretion have been successful at reproducing the broad characteristics of the terrestrial planets in our Solar System \citep{Wetherill:1994,Chambers:2001,Raymond.etal:2004,Raymond.etal:2006a,Obrien.etal:2006,Raymond.etal:2009,Quintana.etal:2002,Quintana.Lissauer:2006,Quintana.etal:2010,Quintana.Lissauer:2014}. These results were all based on a small (up to a dozen) number of realizations performed for each set of initial conditions. A larger set of 50 simulations of planet formation around the Sun was recently performed by \citet{Fischer.Ciesla:2014} which, using the perfect-accretion model, demonstrated the need for a larger suite of simulations in order to infer results from a distribution of final planet configurations.

Several recent developments have opened the door to more sophisticated $N$-body simulations of late-stage planet formation. \citet{Leinhardt.Stewart:2012} and \citet{Stewart.Leinhardt:2012} developed a state-of-the-art collisions model for gravity-dominated bodies that essentially maps collision outcomes based on the masses, collision speed and impact angle for a given two-body collision. This collision model has been implemented into an $N$-body tree code \citep{Bonsor.etal:2015,Lines.etal:2014} and used to examine planetesimal formation.

\citet{Chambers:2013} implemented this collision model into the widely-used $\emph{Mercury}$ integration package \citep{Chambers:2001}. Eight simulations of planet formation around the Sun using this new model were presented and compared to eight simulations that were previously performed using the standard perfect-accretion version of $\emph{Mercury}$. The final planets that formed in each of these sets were shown to be comparable despite the significant difference in the number and frequency of collisions. In addition, the accretion timescales were about twice as long when fragmentation was included.

\citet{Quintana.Lissauer:2014} recently examined planet growth around the Sun with Jupiter and Saturn perturbing the system using the perfect-accretion version of $\emph{Mercury}$. From 2 -- 4 terrestrial planets formed within $\sim$200 Myr in each of the six simulations performed. In this article we expand on this work and perform simulations using the same initial conditions for the Sun-Jupiter-Saturn system as \citet{Quintana.Lissauer:2014}, but we now use the modified version of $\emph{Mercury}$ that allows fragmentation. We have performed an order-of-magnitude larger number of simulations than \citet{Quintana.Lissauer:2014} and \citet{Chambers:2013}. Specifically, 140 simulations using the new fragmentation model and 140 using the perfect-accretion model were performed. This large set of simulations allows us to examine the effects of fragmentation in $N$-body models while accounting for the chaotic nature of these systems, thereby improving upon the two main challenges of $N$-body planet formation models. 

The results from our 280 simulations are presented here. In order to place our simulations in context with the Earth's history, we first present a chronology of Solar System formation in Section~\ref{sec:timescales}. Our numerical model and initial conditions are given in Section~\ref{sec:model}. In Section~\ref{sec:sims} we present results on the effects of fragmentation. We then focus on the Earth-analogs that form and examine the giant impacts onto these planets in Section~\ref{sec:earths}, and a summary and conclusions are given in Section~\ref{sec:conclusions}

\section{Chronology}\label{sec:timescales}

\begin{figure*}
\plotone{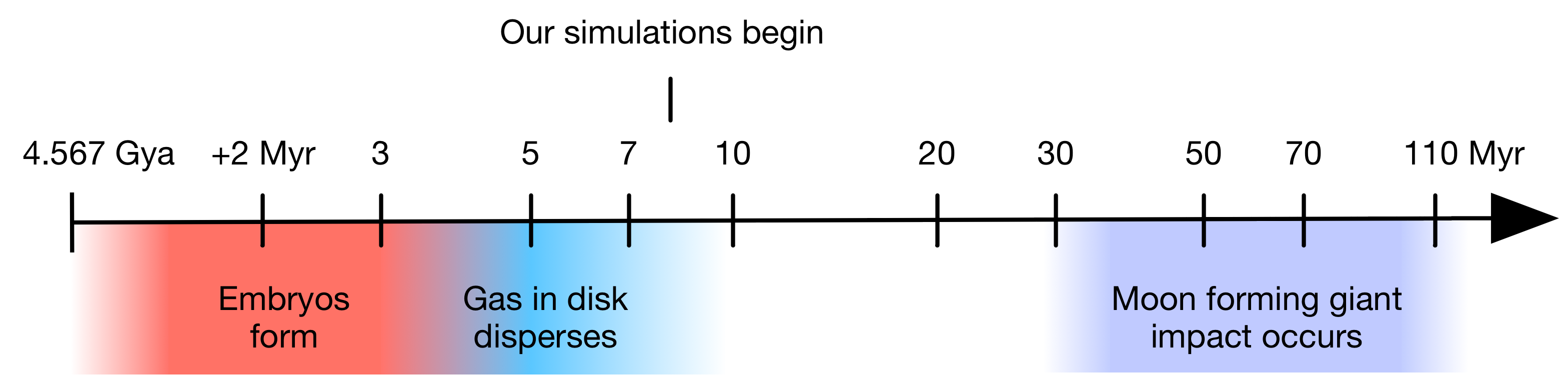}
\caption{An approximate timeline for the formation of the Earth based on geochemical evidence and theoretical models. The oldest known chondrite meteorites date the beginning of Solar System formation to 4.567 billion years ago \citep{Connelly:2012}. Within about 5 Myr, solid material in the protoplanetary disk accreted to form Mars-sized embryos embedded in a swarm of Moon-size and smaller planetesimals. Gas was still present up until about 3 -- 7 Myr \citep{Haisch:2001,Hernandez:2007,Hernandez:2010} during which the cores of the giant planets accreted their gaseous envelopes \citep{Lissauer:1987,Lissauer:2009}. Our simulations start at an epoch at which the gas in the disk has been dispersed and a disk of solid Moon-to-Mars-sized bodies remains. The so-called giant impact phase of planet formation started some time after about 30 Myr, during which the Earth accreted the bulk of its mass. The Moon-forming giant impact also occurred some time after $\sim$30 Myr, but the timeline and impact scenario are still under debate. \label{fig:timeline}}
\end{figure*}

An approximate timeline that places our $N$-body simulations of late stage accretion in context with some of the significant events in the Earth's history is shown in  Figure~\ref{fig:timeline}. The beginning of the Solar System's formation is often defined as the time that the first solids condensed in the Sun's protoplanetary disk as it cooled. The oldest known substances are Calcium Aluminum Inclusions (CAIs) found in carbonaceous chondrite meteorites, and radiometric dating places their origins at $\sim$4.567 billion years ago \citep{Connelly:2012, MacPherson:2014}. 

Planets then grow by a pair-wise accretion process from a protoplanetary disk of gas and dust that is a remnant of the star's formation \citep{Safronov:1969, Lissauer:1993}. The early stages of planet formation, from dust to km-sized objects, remain poorly understood, but theories suggest that growth may occur via a gravitational sticking process  \citep{Weidenschilling:1977, Krijt:2015} or by gravitational instabilities \citep{Goldreich:1973} to create km-sized planetesimals. In the next stage, a runaway growth phase occurs where the largest bodies accrete material faster than smaller bodies until they are large and massive enough for orbital repulsion to take over \citep{Kokubo.Ida:2000}. An oligarchic growth phase then takes place, yielding a disk with approximately Mars-sized objects embedded in a swarm of Moon-sized and smaller bodies \citep{Kokubo.Ida:1998}. 

Geochemical evidence from Martian meteorites provides constraints on the formation timescale of Mars. In turn, these constraints can be used as a proxy for the formation timescale of Mars-sized planetary embryos given that Mars is thought to be essentially a planetary embryo that escaped a significant amount of accretion or erosion during the final stages of planet formation \citep{Dauphas:2011}. Isotopic evidence suggests that Mars grew to half its size within about 1.2 -- 2 Myr and was mostly formed (accreting $\sim$90\% of its mass) within about 5 Myr \citep{Dauphas:2011,Tang:2014}. This would place constraints on the formation of Mars-sized embryos in the protoplanetary disk to within about 5 Myr from the start of Solar System formation. The lifetime of the gas component in the disk is thought to have been $\sim$3 -- 7 Myr \citep{Haisch:2001,Hernandez:2007,Hernandez:2010}, during which the gas giant planets like Jupiter and Saturn must have formed in order to accrete their gaseous envelopes \citep{Lissauer:1987,Lissauer:2009}. Our simulations begin at the epoch at which gas has been dispersed from the disk, the giant planets have formed, and Mars-sized embryos and smaller planetesimals remain and are subject to purely gravitational (no gas drag) perturbations and collisions. 

Isotopic constraints and dynamical simulations place the subsequent formation of the Earth somewhere between 30 -- 100 Myr after the birth of the Solar System \citep{Chambers:2001,Morbidelli.etal:2012}. The Moon is thought to have formed via a giant impact of a large body with the proto-Earth, although the scenario and time of this impact is still under debate. Several different theories have been developed to explain the characteristics of the current Earth-Moon system given the constraints provided by lunar samples \citep{Taylor:1975} and meteorites. The time of the Moon-forming impact is thought to have occurred as early as $\sim$30 Myr and as late as 110 Myr \citep{Canup.Asphaug:2001, Canup:2004,Canup:2008,Halliday:2008,Canup:2012, Cuk.Stewart:2012}, although the upper estimate remains relatively unconstrained by geochronology due to the wide range of ages for lunar rocks \citep{Tanton:2012}. This timeline is applicable for the Solar System and, given that accretion is a chaotic process, is not expected to be the only model for other planetary systems around Sun-like stars. 

In this article we explore the late stage accretion of terrestrial planets around a Sun-like star in a statistical manner in order to understand the types of collisions that lead to planet growth and the final giant impacts that bombard the planets that form. We can then compare the distributions of planet characteristics to those of the Solar System and estimate how common systems like our terrestrial planets may be around other stars in our galaxy.

\section{Numerical model} \label{sec:model}
The late stages of terrestrial planet formation -- from embryos and planetesimals to planets -- have been studied extensively with $N$-body algorithms that have greatly improved in both speed and functionality over the past few decades \citep{Wetherill:1994, Wetherill:1996, Chambers:1999, Chambers:2013,Levison:2000}. Early on, the primary goal of many of these studies was to broadly reproduce the masses and architecture of the four terrestrial planets in our Solar System. The wide diversity of newly discovered exoplanetary systems motivated the modification of these $N$-body models to address planet formation in different environments, such as in multiple-star systems \citep{Chambers.etal:2002, Quintana.etal:2002, Quintana.etal:2007, Quintana.Lissauer:2006}, around stars of various stellar types \citep{Lissauer:2007,Raymond.etal:2007b}, in the presence of inward migration of disk material \citep{Mandell.etal:2007} and in the presence of various giant planet companions \citep{Raymond.etal:2005a,Raymond:2006,Quintana.Lissauer:2014}. Models that included the accretion of water and volatile compounds to explore how Earth became a habitable planet soon followed \citep{Raymond.etal:2004}, as did higher resolution (a larger number of smaller bodies) simulations \citep{Raymond:2006, Obrien.etal:2006}. 

While a simple perfect-accretion model allows a broad study of the numbers and types of planets that can form under various circumstances, a model that allows fragmentation as an outcome of collisions is necessary to explore the history of collisions that lead to the formation of Earth-like planets and their affect on the final planet's composition, dynamics and potential habitability. \citet{Leinhardt.Stewart:2012} developed a self-consistent set of scaling laws to map out the dynamical outcomes of gravity-dominated two-body collisions based on the masses, velocities and geometry of the bodies involved. \citet{Chambers:2013} implemented this collision mapping model into the $N$-body integration package $Mercury$, which we adopt for the simulations presented herein. The collision model is described next, followed by a description of the initial conditions of our simulations.

\subsection{Collision model} \label{sec:collisions}

For each collision event throughout an $N$-body simulation, the specific impact energy in the center of mass reference frame, $Q$, can be computed by 

\begin{equation}
\label{equ:eq1}
 Q  = \frac{\mu \; \vimp^2}{2 \; (\mimp + \mtar)}
\end{equation}

where $\mimp$ is the mass of the impactor, $\mtar$ is the mass of the target, $\mu$ is the reduced mass, where 
\begin{equation}
\label{equ:eqmu}
\mu = \frac{\mimp \; \mtar}{\mimp + \mtar} ,
\end{equation}
and $\vimp$ is the impact velocity.

\citet{Leinhardt.Stewart:2012} performed a large number of high-resolution two-body collision simulations to explore the outcomes using a wide range of initial conditions, including impactor mass ($\mimp$), target mass ($\mtar$), impact angle $\theta$ and impact velocity $\vimp$. By comparing $Q$ with the criteria for catastrophic disruption between two colliding bodies, $Q^*$ (which by definition is the specific impact energy needed to disrupt half of the total mass), and evaluating the outcomes for all of their collision simulations, they derived analytic equations to map out the transitions between collision regimes. For a given collision, this model provides the outcome (for example, a grazing impact with partial erosion) as well as the size and velocity distributions of the post-impact bodies.

\citet{Chambers:2013} implemented these collision outcomes into $Mercury$ and introduced a new input parameter $m_{frag}$ that defines a minimum fragment mass for all collisions. In this modified version of $Mercury$, if a collision leads to fragmentation, the mass of the largest remnant remaining from the target, $m_{lr}$, is recorded. The remaining mass from the target (or the impactor in some scenarios), $m_{fragtot}$, is broken up into $x$ equal-sized fragments, where $x$ = $m_{fragtot}$ / $m_{frag}$ (with any remainder mass less than $m_{frag}$ left as is). The fragments are gravitationally dispersed at slightly above the escape velocity in uniform directions within a plane that includes the center-of-mass of the system. The evolution of these new fragments is then followed from that time forward while they gravitationally interact with other bodies in the disk. Note that the collision model of \citet{Leinhardt.Stewart:2012} does not describe the fragments as a set of equal size bodies. The simplification of equal size fragments implemented in $Mercury$ is intended to keep simulations computationally tractable, as the true fragments would have a size distribution with many smaller fragments.

The outcome regimes of two-body collisions in our simulations with fragmentation include:

\begin{enumerate}

\item A collision with the central star (or outer giant planet) in which case the body is removed from the simulation

\item Ejection from the system if a body orbits beyond 100 AU from the central star

\item Perfect accretion (inelastic collisions) where the mass of the new body $m_{lr}$ is equal to \mimp + \mtar

\item A grazing collision that can lead to accretion or erosion. Grazing collisions occur when less than half of the projectile interacts with the target. In this case the impact parameter $b$ is greater than the critical impact parameter $b_{crit}$ = $r_{tar}$ / ($r_{tar}$ + $r_{imp}$), where $r_{tar}$ is the radius of the target and $r_{imp}$ is the radius of the impactor

\item A non-grazing collision, where $b$ $<$ $r_{tar}$, that can lead to accretion or erosion

\item A `hit-and-run' collision, discussed in detail in \citet{Asphaug.etal:2006,Genda:2012}, in which the target does not gain or lose a significant amount of mass, but the impactor may suffer fragmentation, so $m_{lr}$ = \mtar and $m_{fragtot}$ $\le$ \mimp

\end{enumerate}

At each collision event the initial and final masses, positions and velocities are recorded for the target, impactor and any fragments. In addition, parameters to characterize the collision are output such as the fate (type of collision), impact angle and impact velocity.  All of the information needed to evaluate the collision history for planets that form in an $N$-body simulation is made available in the output of $Mercury$.

\subsection{Initial conditions}\label{sec:ic}

Our simulations begin at the epoch of planet formation in which runaway and oligarchic growth have already taken place \citep{Kokubo.Ida:1998} resulting in a bimodal disk mass distribution around the Sun. In total, 26 embryos and 260 planetesimals reside in our initial disk. The embryos are approximately Mars-sized ($r_{em}$ = 0.56 \rearth $\approx$ 3500 km; $m_{em}$ = 0.093\mearth) and are spread among approximately Moon-sized planetesimals ($r_{em}$ = 0.26\rearth $\approx$ 1600 km; $m_{em}$ = 0.0093 \mearth). All bodies are assumed to have a density of 3 \gcm to determine their radii and the bodies are placed between 0.35 AU to 4 AU from the Sun. The surface density of solids varies as $a^{-3/2}$, consistent with Solar Nebula models \citep{Weidenschilling:1977}, yielding a total disk mass of 4.85 \mearth.  The bodies begin on nearly circular and coplanar orbits with all other orbital elements (argument of periastron, longitude of ascending node and mean anomaly) selected at random. At the start of our simulations we include a Jupiter-mass planet at 5.2 AU and a Saturn-mass planet at 9.6 AU from the Sun at their present orbits and assume that these giant planets have already formed and that any gas in the disk has been dispersed.  The problem is then purely gravitational and collisional, and any damping of the eccentricities and inclinations of embryos arises from dynamical friction from the swarm of smaller bodies and not by gas drag.

We implemented the same initial disk model as was used by \citet{Quintana.Lissauer:2014}, in which several dozen simulations were performed with the perfect accretion model to examine the role of Jupiter and Saturn on water delivery to the terrestrial planets. This disk model was adapted from \citet{Chambers:2001} (which was also used by \citet{Chambers:2013} to test the fragmentation model), but we extrapolated our disk out from 0.35 to 4 AU whereas the disk in \citet{Chambers:2001, Chambers:2013} was truncated at 2 AU.  The sizes of embryos and planetesimals remained the same in all studies, but there were only 14 embryos and 140 planetesimals in \citet{Chambers:2001} compared to our 26 embryos and 260 planetesimals. 

Because $N$-body simulations are highly stochastic, we examine a large number of simulations in order to interpret our results statistically. We performed 280 $N$-body simulations -- 140 with fragmentation and 140 without -- of accretion from the disk of bodies around a Sun-like star with Jupiter and Saturn analogs. The same set of elements for all bodies in the initial disk were used in each simulation apart from minuscule variations. Starting with the second simulation, we iteratively displaced a planetesimal near 1 AU by 1 meter along its orbit thereby creating a set of 140 initial disks that were all (slightly) different. In each simulation the evolution of all bodies in the disk was followed for 1 Gyr (note that in Section~\ref{sec:earths} we extended the fragmentation simulations to 2 Gyr).

An additional parameter that is needed for the integration is the minimum fragmentation mass which ultimately constrains the total number of bodies in a given simulation.  We set the minimum fragmentation mass to be 0.0047 \mearth, which is $r_{frag}$ = 0.2 \rearth $\approx$ 1300 km, which is a little under half a lunar-mass.  This is the same value used in \citet{Chambers:2013} and was adopted here for comparison. While smaller minimum fragmentation mass values would allow a better representation of the protoplanetary disk that formed our Solar System planets, this choice allowed us to constrain the total number of bodies in order to perform a large number of simulations. With this choice of $m_{frag}$, the average number of fragments created in our simulations was 270, comparable to the number of initial embryos and planetesimals. The sensitivity of the results to this parameter are discussed in Section~\ref{sec:limits}.

\section{Simulation results}\label{sec:sims}
In this section we present the results from our 280 simulations of planet formation around the Sun with Jupiter and Saturn analogs perturbing the system from their current orbits.  We ran 140 simulations of this system using the new fragmentation model and 140 simulations using our standard (perfect accretion) model in order to quantify the differences resulting from fragmentation. 

\begin{figure}
\plotone{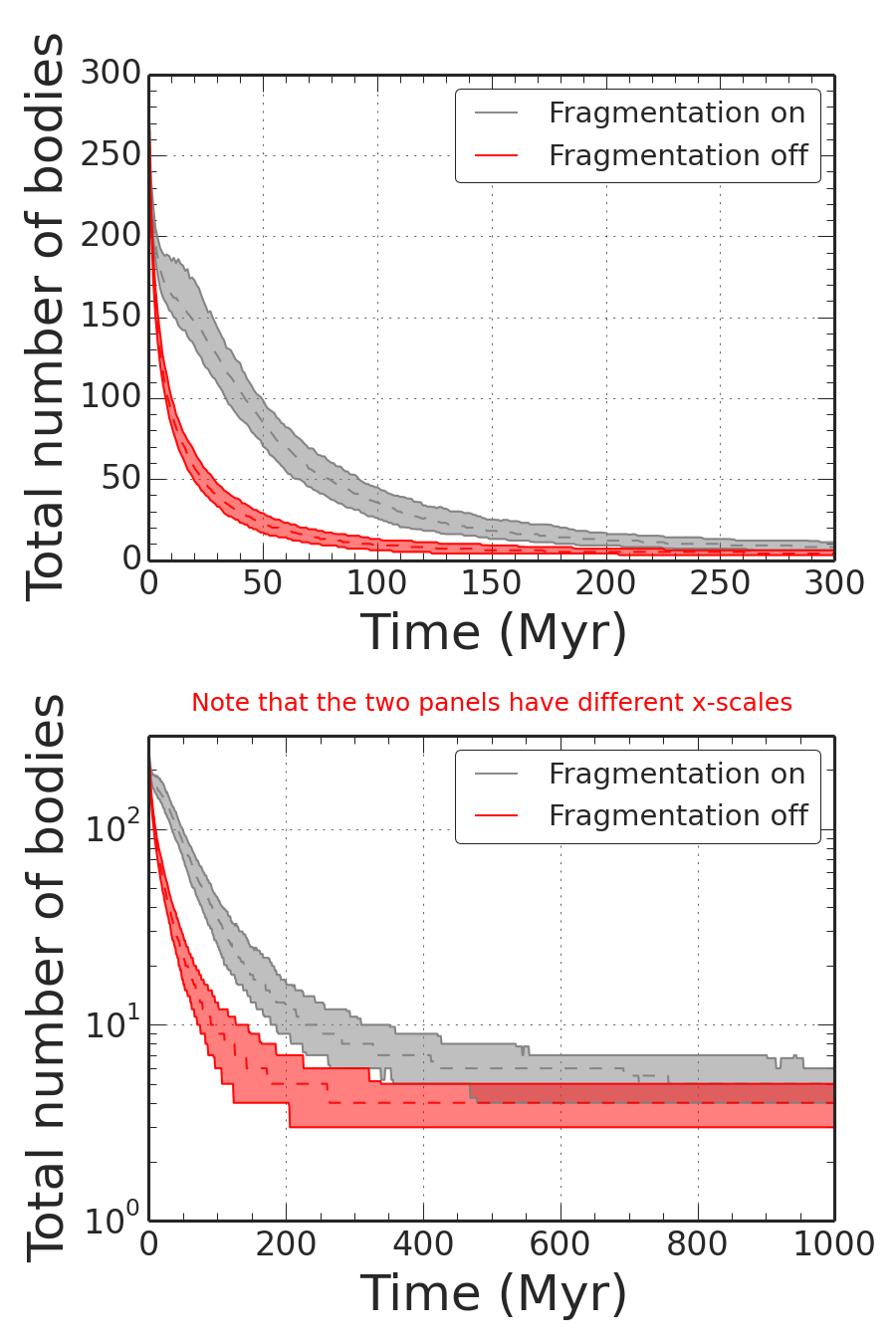}
\caption{The total number of bodies remaining in our simulations as a function of time. The shaded regions show the 15th and 84th percentile which is equivalent to the 1-$\sigma$ bounds of the distribution. The red region shows the number of bodies in our simulations that utilize the standard $\it{Mercury}$ $N$-body integrator while the grey region shows the results from simulations that include fragmentation. All simulations shown in this figure begin with 26 Mars-sized embryos embedded in a disk of 260 Moon-sized bodies and include Jupiter and Saturn analog planets. The upper and lower panels show the same data but with a linear and log vertical scale, respectively, to allow a better view of when the number of planets finalize. It should be noted that the horizontal scales of the two panels differ. Although both sets produce final planetary systems that are consistent, the collisional evolution is quite different among the two sets. \label{fig:SJS_nbodies_time}}
\end{figure}

Figure~\ref{fig:SJS_nbodies_time} shows the total number of bodies (1-$\sigma$ ranges are shaded) from 140 simulations as a function of simulated time, with the standard simulations (without fragmentation) shown in red and those with fragmentation enabled shown in grey. All simulations began with the same 289 bodies: 26 Mars-sized embryos, 260 Moon-sized planetesimals, Jupiter, Saturn, and the Sun.

When fragmentation is enabled, the total number of bodies varies as the system undergoes accretion and fragmentation. On average, the total number of bodies drops at a much slower rate compared to simulations that assume perfect accretion, as expected, and for the first few hundred million years there is a significant difference in the number of bodies at any given time among the two sets. With fragmentation, the number of bodies drops to half the initial number in 23 Myr whereas the perfect accretion simulations see the number of bodies fall to half the initial value in just 4 Myr -- a factor of 6 difference. This leads to longer accretion timescales when fragmentation is included, although this divergence in accretion timescales is primarily due to the larger number of small bodies. The average time for the final planets to form (and for most of the small bodies to be accreted or ejected) with fragmentation is about 200--300 Myr, in contrast to the non-fragmentation simulations where accretion was complete by 100--200 Myr. Despite the different collision histories and timescales, the simulations ultimately converge towards a comparable final number of planets (as shown in the lower panel of Figure~\ref{fig:SJS_nbodies_time} which shows the same data in log-space), in agreement with Chambers (2013).

\begin{figure}
\plotone{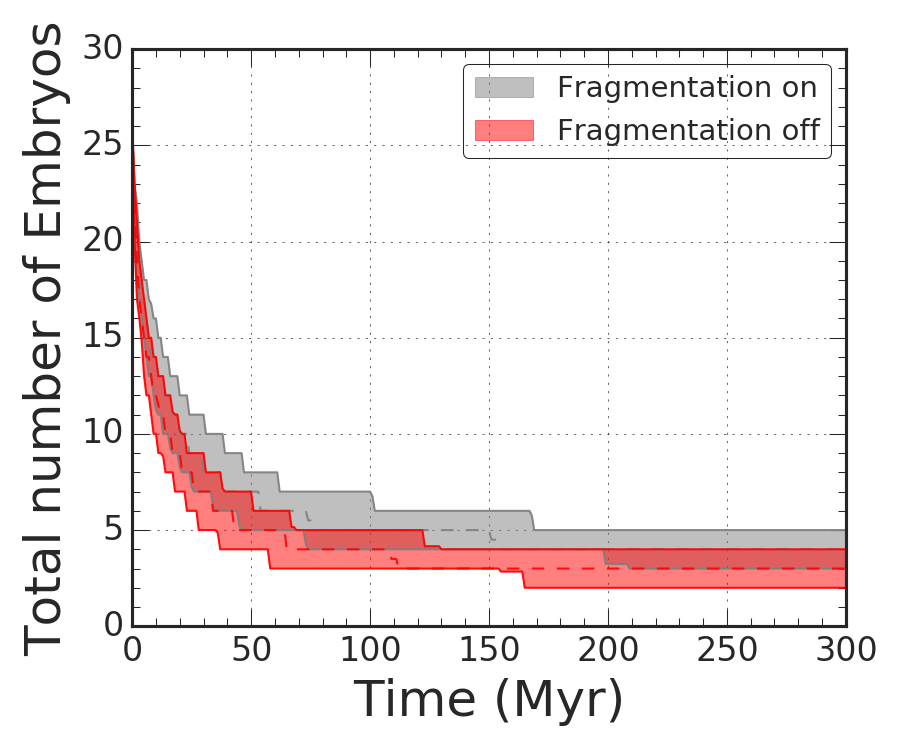}
\caption{The total number of embryos remaining in our simulations as a function of time, where each initial disk begins with 26 Mars-sized embryos (as well as 260 planetesimals not included here). The symbols and colors are the same as those described in the caption of Figure~\ref{fig:SJS_nbodies_time}.  The number of embryos in the fragmentation model decreases at a slower rate than in the standard (perfect accretion) model, but not significantly so. 
\label{fig:SJS_embryos_time}}
\end{figure}

Figure~\ref{fig:SJS_embryos_time} compares just the number of embryos -- which are the initial 26 Mars-sized bodies -- as a function of simulation time. The number of embryos over time does not differ for the two different models as dramatically as the total number of bodies, thus the main difference among the two models shown in Figure~\ref{fig:SJS_nbodies_time} can be attributed to the planetesimals and fragments. On average 3.9$\pm$0.9 final planets formed in the simulations with fragmentation while 3.0$\pm$0.8 planets formed in the simulations with perfect accretion. In the 8 fragmentation runs performed in \citet{Chambers:2013}, the average number of final planets was 4.4$\pm$0.9, consistent with our findings. This demonstrates that despite the different formation histories of the planetary systems, fragmentation does not have a strong impact on the number of final planets.

\begin{figure}
\plotone{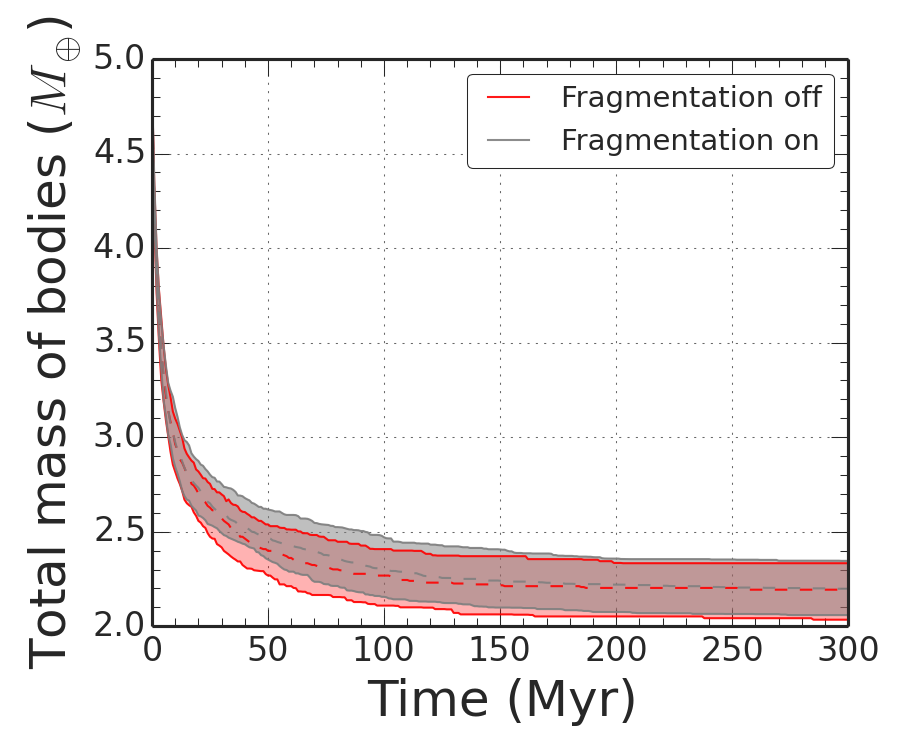}
\caption{The median of the total mass remaining in the simulations as a function of time.  The symbols and colors are the same as those described in the caption of Figure~\ref{fig:SJS_nbodies_time}. Both models produce comparable results in terms of the total mass in the system remaining as a function of time. This is in contrast to the total number of bodies which decreases more slowly in the fragmentation model. This indicates that much of the material that is fragmented is re-accreted at later times rather than ejected from the system.
\label{fig:SJS_mass_time}}
\end{figure}

The evolution of the average total mass remaining in the simulations is shown in Figure~\ref{fig:SJS_mass_time}. During the first few hundred million years there is on average marginally more mass available for collisions at a given time when fragmentation is included. By the time the final planets have formed, there is considerable overlap among the two sets of simulations.  This is an indicator that a large fraction of the mass that is fragmented is ultimately re-accreted before it can escape the system.

\begin{figure}
\plotone{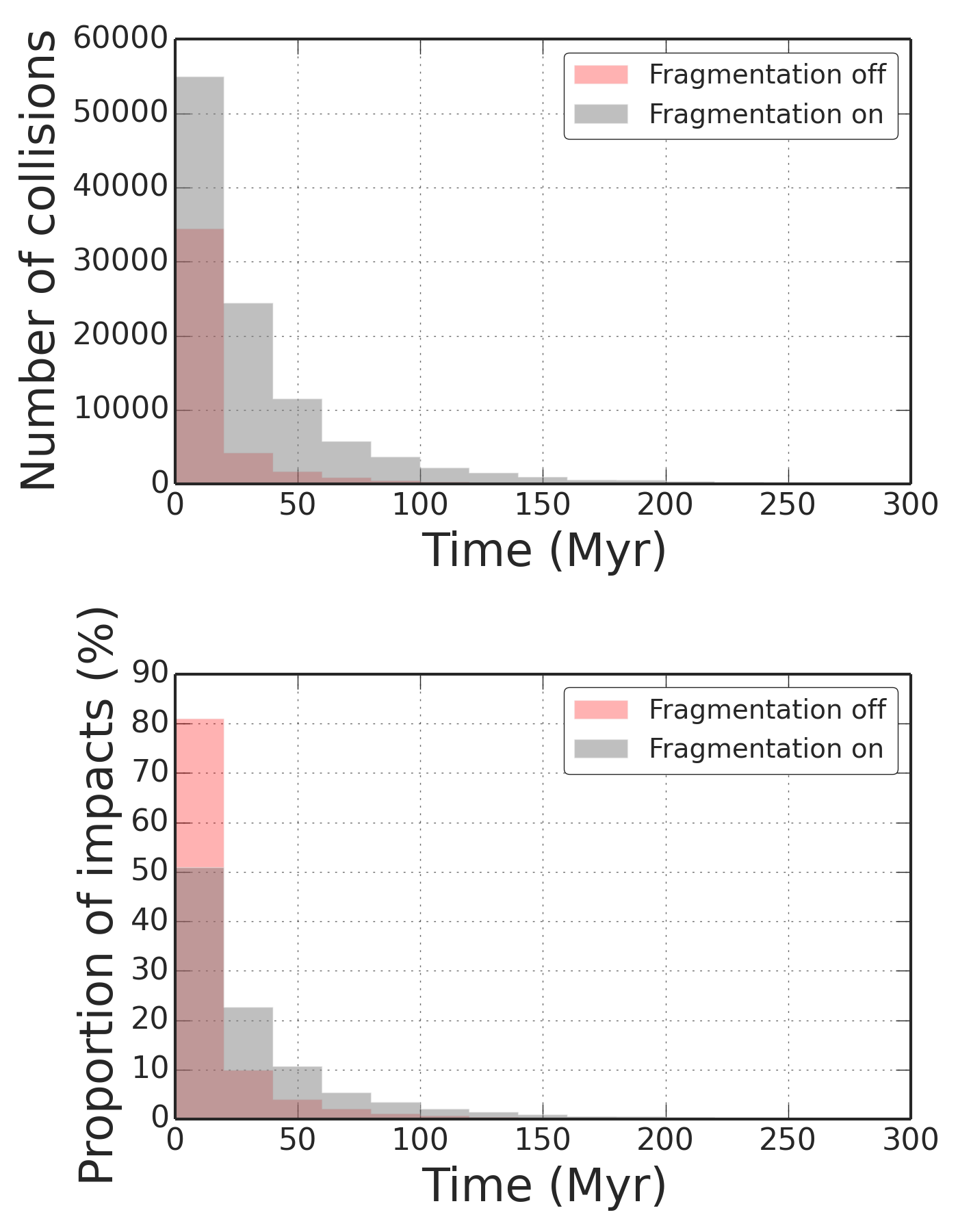}
\caption{The number of impacts as a function of simulation time with fragmentation (gray) and without (red). The top panel shows the total number of impacts as a function of time for all our simulations while the lower panel shows the percentage of total impacts falling within that 20 Myr bin. While the total number of collisions is higher when fragmentation is included, only 50\% of the collisions occur in the first 20 Myr, whereas 80\% of the collisions occur in the first 20 Myr in the perfect accretion model. By 50 Myr, accretion is mostly complete when fragmentation is not included. Collisions in the fragmentation simulations continue for an additional 100 Myr.  Despite these differences, the final planets that form in each set are comparable in terms of final mass and number.
\label{fig:impact_rate}}
\end{figure}

The number of collisions as a function of simulation time is shown in Figure~\ref{fig:impact_rate} and includes all collision events that occur in the 140 simulations with fragmentation enabled (shaded gray bars) and the 140 simulations that assumed perfect accretion (red shaded bars). Both the total number of collisions (top panel) and the percentage of total collisions (lower panel) are quite different among the two sets within the first 20 Myr. The total number of collisions is significantly higher when fragmentation is included, as expected. While  there are 60\% more collisions within the first 20 Myr using the fragmentation model, these collisions only contribute to about 50\% of the total number of collisions that occur in this model, whereas nearly 80\% of the collisions in the standard model occur within the first 20 Myr. The time for the number of collisions to drop by a factor of two (the collisional half-life) for the perfect accretion simulations is 4 Myr compared with 19 Myr when fragmentation is included. The average time it takes for 90\% of collisions to occur is 37 Myr for the perfect accretion simulations compared to 82 Myr for the fragmentation simulations, showing that the accretion timescales are approximately doubled when fragmentation is included.

In this section we have demonstrated that during the first 200 Myr or so the results from our fragmentation model differ significantly from the perfect accretion paradigm. In the next section we use this fragmentation model and focus on giant impacts onto Earth-like worlds to better understand the frequency and timescales of events that influence the final planet properties.
 
\section{Giant impacts onto Earth-analogs}\label{sec:earths}
We now focus on Earth-like planets that form in simulations using our fragmentation model to evaluate characteristics of these planets that depend on their collision history. The Earth's geological record has established that a large number of significant impacts onto Earth helped to shape its composition and habitability.  Small-scale examples include cratering events, some of which have caused mass extinctions \citep{Lissauer.dePater:2013}. Larger-scale events include the Moon-forming impact and other giant impacts that have affected the Earth on a global scale. In our Solar System collisions ultimately led to the formation of four terrestrial planets and three moons within 2 AU of our Sun, with at least one of those planets remaining habitable despite the multitude of impacts.

It is unclear, however, if conditions in the Solar System were special for the Earth to form and retain an atmosphere and oceans. In theory, giant impacts can deliver or strip away water and other volatiles that are necessary ingredients for carbon-based life \citep{Cameron:1983, Ahrens:1993, Lissauer.dePater:2013}. After an impact it can take $10^5$-$10^6$ years for a planet to cool \citep{lupu14}. The time of the final giant impact is also important because, for example, if a giant atmosphere-stripping impact occurs early enough in the planet's accretion history, there may still be volatile-rich material available in the protoplanetary disk to accrete. If these types of impacts occur late when most of the solid material in the disk has been accreted or ejected from the system, then the planet may have to rely on distant water/volatile-rich material like comets (which have a low collision rate with bodies in the terrestrial region) or else remain depleted of volatiles.

We extended the fragmentation simulations presented in the previous section from 1 to 2 Gyr to examine the collisions that led to the formation of Earth-analogs on longer timescales. We define an Earth-analog as a final planet with a mass greater than 0.5 \mearth that orbits with a semimajor axis between 0.75 and 1.5 AU from the Sun. Of the 140 simulations that included fragmentation, 96 simulations produced one Earth-analog, 34 produced two, and 10 simulations resulted in zero Earth-analogs.

\begin{figure}
\plotone{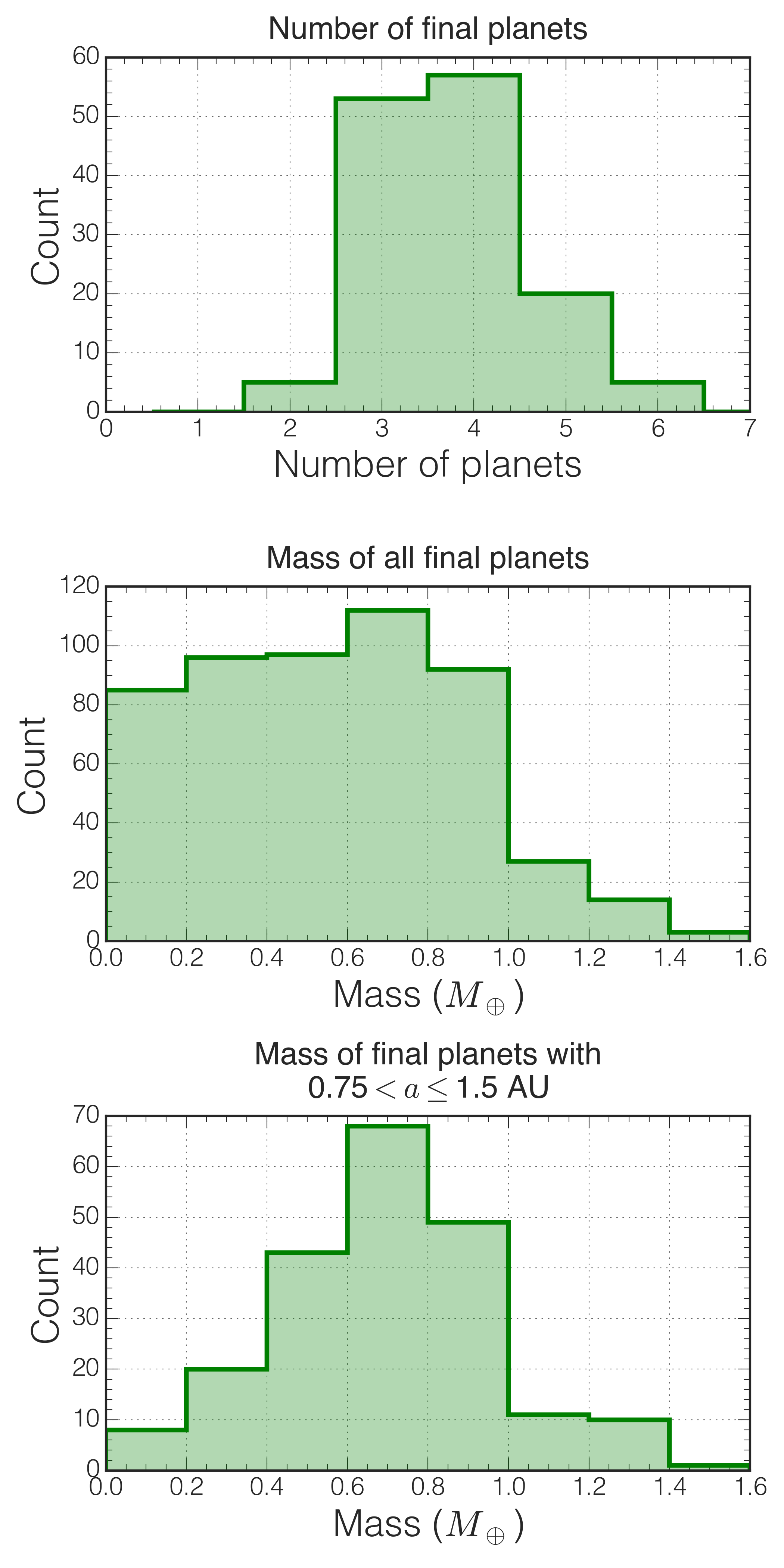}
\caption{Histograms of the final planets formed in our simulations that included fragmentation. The top panel is the total number of final planets for all fragmentation simulations, where a planet is defined as a body with a mass larger than 0.9 Mars-mass. The middle panel shows the distribution of the masses of all final planets. The lower panel shows the mass distribution of final planets with final semimajor axes between 0.75 and 1.25 AU, which is the range we define for an Earth-analog. 
\label{fig:numPlanetsHist}}
\end{figure}

Figure~\ref{fig:numPlanetsHist} shows distributions of the number of final planets, the mass of the final planets, and the mass of the Earth-analogs that form in our fragmentation simulations. From 2 -- 6 final planets form with masses ranging from that of Mars ($\sim$0.1 \mearth) to 1.6 times the mass of Earth. For the region where Earth now resides (0.75 -- 1.5 AU), fewer low-mass planets form and the distribution is more centered around 0.6 -- 1 \mearth. We note that we also form analogs of Mars, Venus and Mercury, although our choice for the inner edge of the disk (0.35 AU) is not well suited to study the formation of Mercury-analogs. 

\subsection{Quantifying giant impacts} \label{sec:quantify}
To examine the collisions experienced by the Earth-analogs formed in our models, we first determined how to quantify the impacts. \citet{Stewart.etal:2015} performed a suite of 3D hydrodynamic simulations of two-body gravitational collisions to study the types of collisions that could blow off a portion or all of an atmosphere or ocean, assuming the bodies involved had accreted an atmosphere and/or ocean. In their study, when the two bodies collide, the shock waves propagate and affect their atmospheres. Using a wide range of target and impactor masses, impact speeds, and impact angles, \citet{Stewart.etal:2015} found that variations in the shock pressure fields, and therefore the fraction of atmosphere loss, are highly sensitive to the impact parameter and mass ratio of the impactor to the target. Using the results from these hydrodynamic simulations, \citet{Lock.etal:2015} empirically derived a modified version of the specific impact energy $Q$ (as defined in Equation~\ref{equ:eq1}), designated $Q_s$, that better encapsulates variations in the shock pressure field and includes only the interacting mass of the two bodies:

\begin{equation}
\label{equ:eq3}
 Q_s  = Q\prime \left(1 + \frac{\mimp}{\mtar} \right) (1 - b )
\end{equation}

\begin{figure}
\plotone{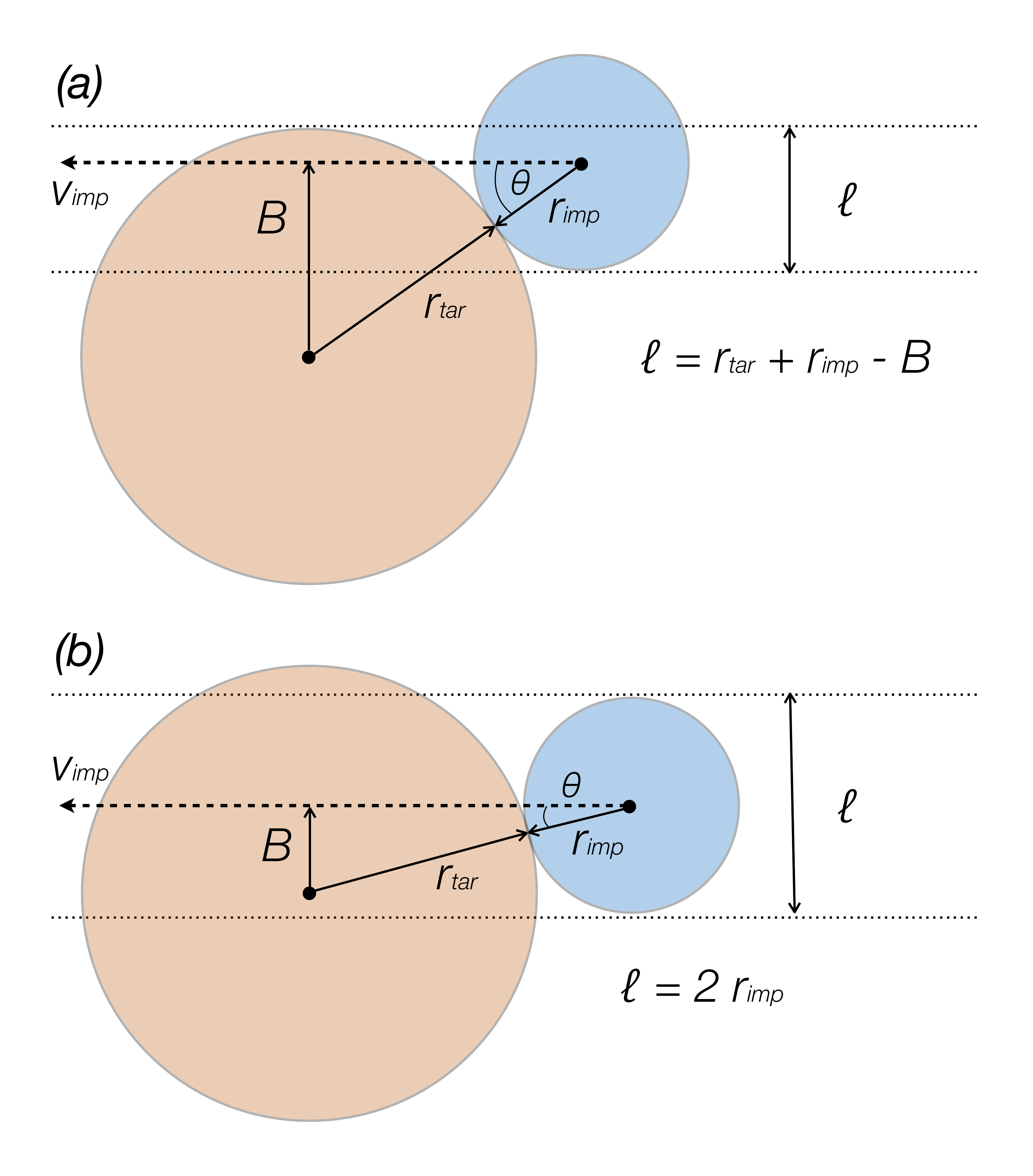}
\caption{Geometry of a two-body collision. The parameter $l$ is the projected length of the interacting mass and takes the value of $\rimp + \rtar - B$ if the impact is grazing (panel a) and $2 \rimp$ if the impactor fully interacts with the target (panel b).
\label{fig:geometry}}
\end{figure}

At the point of contact, assuming the target is stationary, the impactor (the blue body in Figure~\ref{fig:geometry}) is traveling with impact velocity \vimp toward the target body (shown in brown). The impact parameter is defined as $b$ = sin$\theta$ and the projected distance between the center of the two bodies at contact is $B$ = (\rimp + \rtar) $b$, where \rimp and \rtar are the radii of the impactor and the target, respectively. The projected length of the impactor that intersects the target is $l$ and, depending on whether the impactor fully interacts with the target, can take a value of 
\begin{equation}
l = \begin{cases} 2 \; \rimp &\mbox{if } B + \rimp \leq \rtar \\ 
\rimp + \rtar - B & \mbox{if } B + \rimp > \rtar. \end{cases} 
\end{equation}
These two cases are shown in the upper and lower panels of Figure~\ref{fig:geometry}.

The specific impact energy, $Q_s$, is a function of the ratio of the impactor mass ($\mimp$) to the mass of the target ($\mtar$), the impact parameter $b$, and $Q^\prime$ which is a modified version of $Q$ that takes into account only the interacting mass of the two bodies. As shown in \citet{Lock.etal:2015}, $Q^\prime$ can be calculated by scaling $Q$ by the ratio of the reduced mass for the interacting mass, $\mu_\alpha$, to the reduced mass $\mu$ as defined in Equation~\ref{equ:eqmu}
\begin{equation}
\label{equ:eq2}
 Q\prime  = \frac{\mu_\alpha}{\mu} Q
\end{equation}
The interacting reduced mass is 
\begin{equation}
\mu_\alpha = \frac{\alpha \; \mimp \; \mtar}{\alpha\; \mimp + \mtar}
\end{equation}
where $\alpha$ is the fraction of the impactor mass that interacts and is defined as
\begin{equation}
\alpha = \frac{3 \; \rimp \; l^2 - l^3}{4 \; \rimp^3} 
\end{equation}
The original and full derivation of these impact geometries are given in \citet{Leinhardt.Stewart:2012} and \citet{Lock.etal:2015}. 

To associate $Q_s$ to physical events on planets, \citet{Stewart.etal:2015} showed that, for an atmosphere-to-ocean ratio of 1:300, an impact with $Q_s$ $>$ 10$^7$ J/kg was needed to strip 100\% of an atmosphere, and an order of magnitude greater energy ($Q_s$ $>$ 10$^8$ J/kg) was needed to remove an entire ocean. Evaluating $Q_s$ for collisions in previously published $N$-body simulations (ones that assumed perfect accretion), \citet{Stewart.etal:2015} found that impacts that could cause partial atmospheric loss were common but impacts with energies sufficient to blow off an entire atmosphere or ocean were rare. Note the presence of an ocean can significantly enhance the loss of atmosphere during a giant collision due to both evaporation which can push out an atmosphere and a lower shock impedance of the ocean compared to a rocky surface \citep{Genda.etal:2005}. 

Herein, we adopt $Q_s$ as our parameter of interest to evaluate the collisions in our fragmentation simulations, particularly those that lead to the formation of an Earth-analog. Using the \citet{Stewart.etal:2015} empirical relations (that considered the 1:300 atmosphere:ocean case), we chose the value of $Q_s$ that is sufficient to remove about half of a planet's atmosphere to define a `giant impact' threshold. Collisions with $Q_s$ that reach or exceed this value, 2$\times 10^6$ J/kg (equivalent to $\log_{10}{Q_s}=6.3$), are considered a giant impact for our purposes of evaluating the frequency of giant impacts onto the growing terrestrial planets.

As shown in \citet{Stewart.etal:2015}, the values of $Q_s$ for collisions that occur during the late stages of terrestrial planet formation span a wide range that overlap with the estimated $Q_s$ values from various Moon-formation models. Leading theories for the formation of the Moon suggest that a giant collision occurred between the proto-Earth and another object, resulting in a remnant disk from which the Earth's Moon formed \citep{Canup.Asphaug:2001, Canup:2004, Canup:2008,Canup:2012, Cuk.Stewart:2012}. Based on estimates of the age of the lunar melt, this collision likely occurred some time between 30 -- 110 Myr after the Solar System began to form \citep{Taylor:1975,Halliday:2008}, however it could have occurred later (see Section~\ref{sec:timescales}). Nonetheless, the Moon-forming giant impact was likely the final significant giant impact during the Earth's formation.

The full range of $Q_s$ for the canonical Moon-forming impact models is about 0.3$\times$10$^6$ -- 2$\times$10$^6$ J/kg \citep{Canup.Asphaug:2001, Canup:2004}. For models that require higher angular momenta \citep{Canup:2012, Cuk.Stewart:2012}, the range of $Q_s$ is about 2$\times$10$^6$ -- 2.5$\times$10$^7$ J/kg \citep{Stewart.etal:2015}. Our choice of $Q_s$ to define a giant impact therefore corresponds to significant ($\sim$50\%) atmosphere loss and also to the minimum impact energy from the more recent high angular momentum Moon-forming impact models \citep{Canup:2012, Cuk.Stewart:2012}. Our results on the frequency and timing of giant impacts presented herein can be used to compare with models of the Moon's formation and can provide insight to the frequency of these types of impacts during late stage planet formation. However, our results are not meant to predict the abundance of actual moon-forming events beyond our Solar System, as many impacts with energies comparable to the Moon-forming impact may not have enough angular momentum to produce a Moon-forming disk. 

\begin{figure}
\plotone{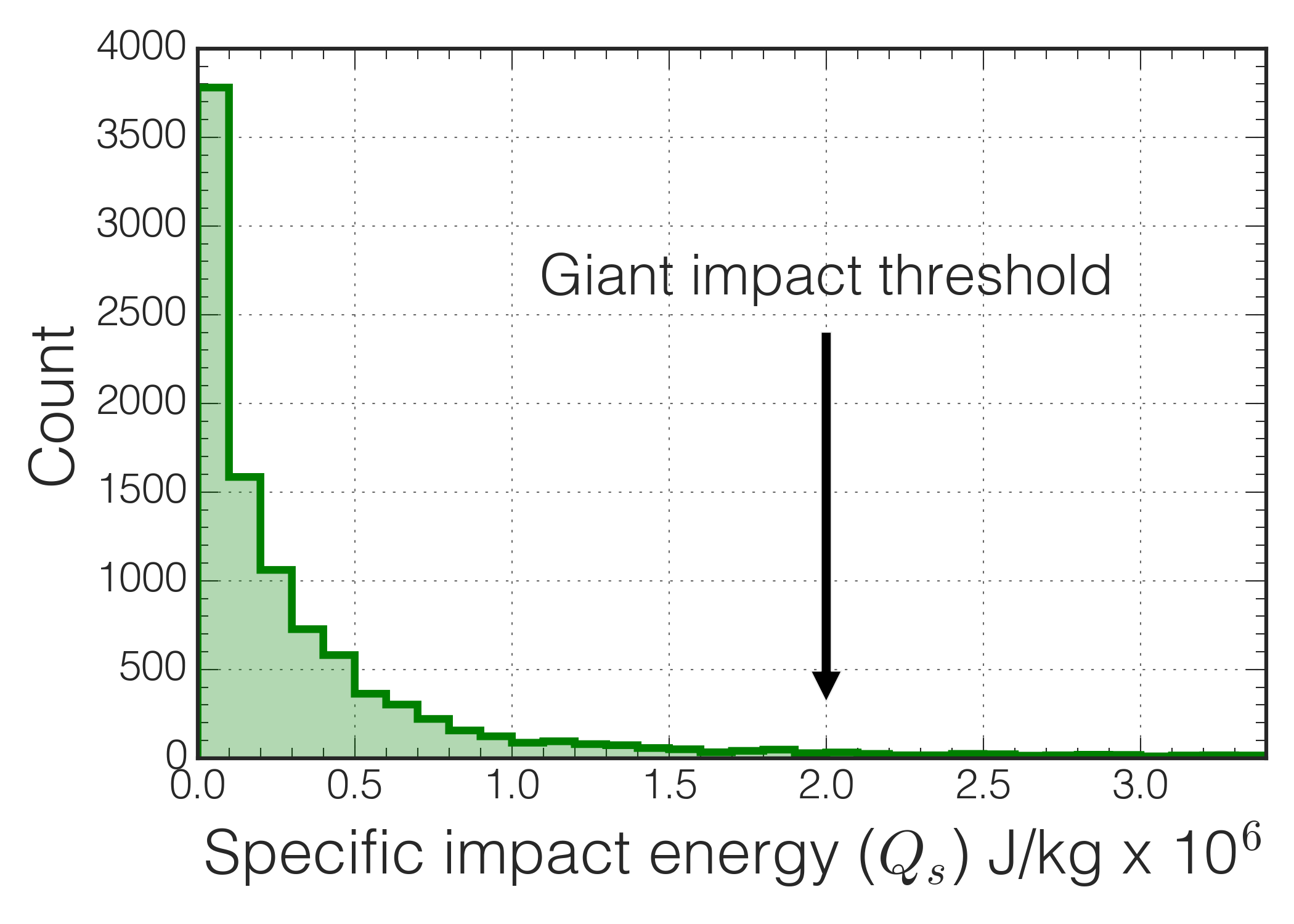}
\caption{The specific impact energy $Q_s$ for all impacts onto Earth-analog planets. An Earth-analog is defined as a planet with mass $>$ 0.5 \mearth that orbits within  0.75 -- 1.5 AU from the Sun. Here we show the energies in linear space to demonstrate that the vast majority of impacts are low energy. We are interested in values of $Q_s$ that reach or exceed $2\times 10^6$ J/kg, a specific impact energy sufficient to strip over half of a planet's atmosphere \citet{Stewart.etal:2015}. This threshold is also comparable to the Moon-forming impact according to recent models \citep{Canup:2012, Cuk.Stewart:2012}. Just 5\% of impacts have energies above our giant impact threshold. 
\label{fig:Qslinear}}
\end{figure}

We computed $Q_s$ for all impacts that were involved in the formation of an Earth-analog for all of our fragmentation simulations. As illustrated in Figure~\ref{fig:Qslinear}, impacts with enough energy to strip entire atmospheres ($Q_s\geq 10^7$ J/kg) and oceans ($Q_s\geq 10^8$ J/kg) are rare, a result consistent with those of \citet{Stewart.etal:2015}. The percentage of impacts with energies comparable to the high angular momentum Moon-forming impact theories is just 5\% of all impacts, while less than 0.2\% of impacts are energetic enough to fully strip an atmosphere of an Earth-analog planet. 

\begin{figure}
\plotone{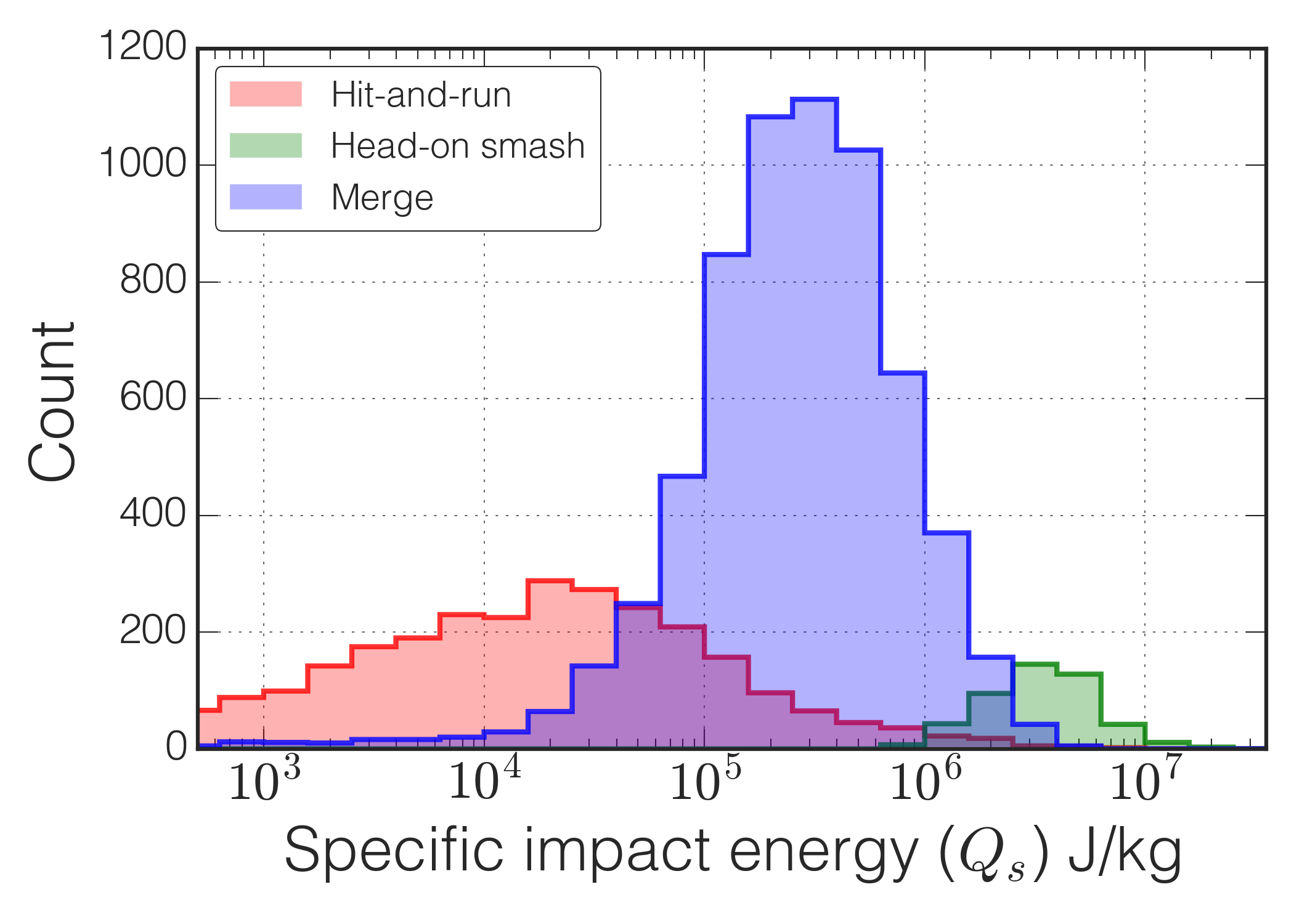}
\caption{Specific impact energies, $Q_s$, for all impact events that led to the formation of an Earth-analog grouped into three types of impacts. Hit-and-run impacts (shaded red) account for 31\% of the impacts that produced Earth-analogs, head-on impacts (shaded green) account for 5\%, and all collisions that led to mergers (shaded blue) account for 64\% of the collisions. Our giant impact threshold is set at $2\times 10^6$ J/Kg, thus the majority of giant impacts onto Earth-analogs in our simulations are head-on collisions.   
\label{fig:qs_fate}}
\end{figure}

Figure~\ref{fig:qs_fate} shows the distribution of these energy values grouped according to three types of impacts: merging events (grazing and non-grazing events that led to net growth), hit-and-run events, and head-on collisions. The majority (64\%) of events result in a merger while 31\% are hit-and-run collisions. The remaining 5\% of impacts are head-on collisions with high values of $Q_s$.  While head-on collision events are uncommon, they account for all events that are capable of stripping a planet of its atmosphere. For collisions that led to the formation of Earth-analogs in the eight fragmentation runs performed by \citet{Chambers:2013}, 70\% resulted in net growth, 30\% were hit-and-run events, and zero collisions led to net erosion. Our results are consistent, considering the relatively small number of simulations and the less massive initial disk used in \citet{Chambers:2013}.

\begin{figure}
\plotone{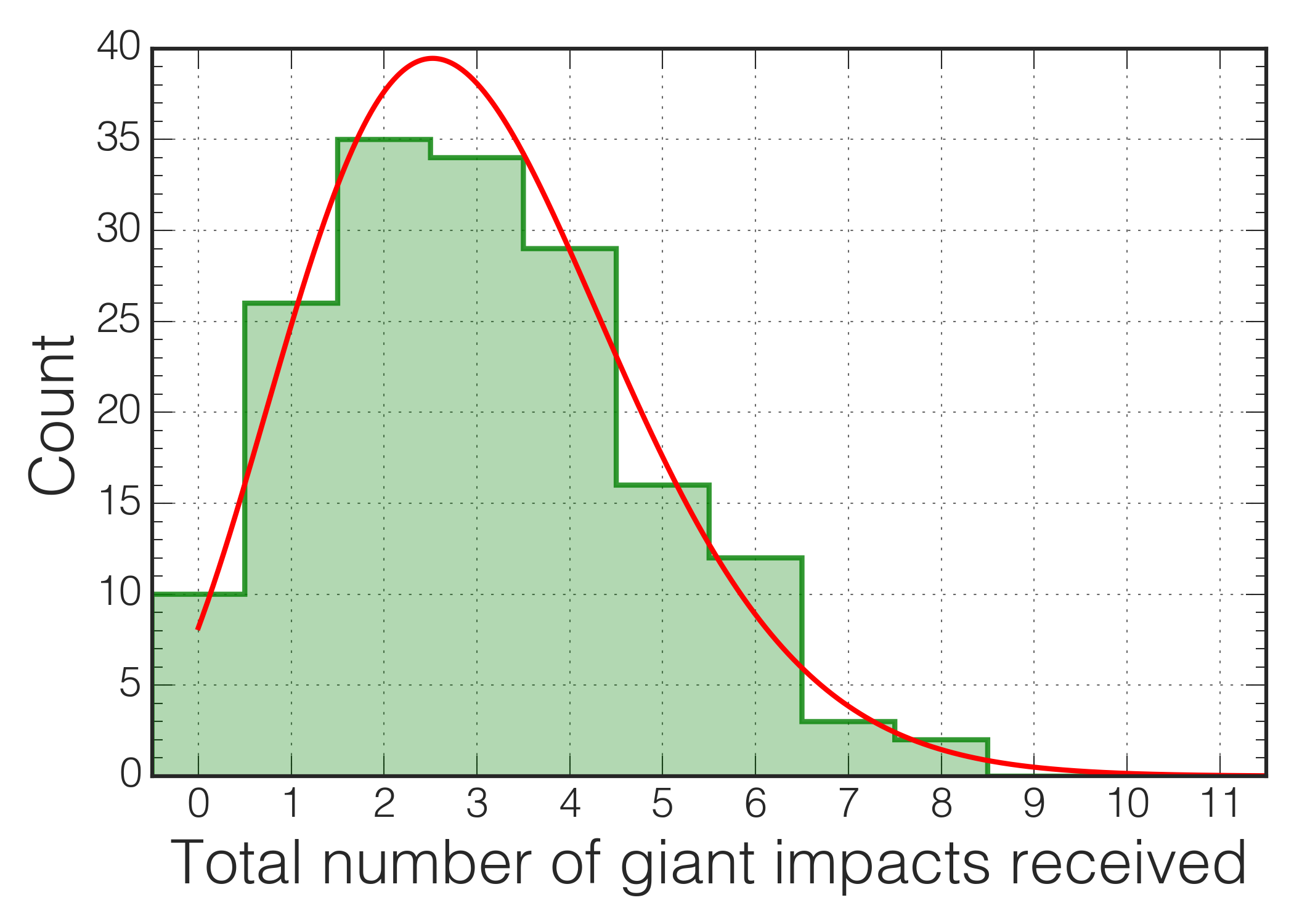}
\caption{A histogram showing the total number of giant impacts received per Earth-analog in all of the 2 Gyr fragmentation simulations. We define a giant impact as an impact where $Q_s > 2\times 10^6$ J kg$^{-1}$. The red curve shows a Poisson probability mass function fit to the data. The mean number of giant impacts per Earth-analog is 3.0 but the number of giant impacts received ranges from 1 -- 8.
\label{fig:numgiantimpacts}}
\end{figure}

We next computed the total number of giant impacts experienced by each Earth-analog during their formation. All but seven of the 167 Earth-analogs experienced at least one giant impact. A histogram of these results is shown in Figure~\ref{fig:numgiantimpacts}, where the bins with the highest occurrence is at two and three giant impacts. Owing to this being a discrete distribution with a number of events occurring within a fixed time window, we are able to describe the resulting probability mass function as a Poisson distribution. We fit these data of the number of giant impacts and found best fitting parameters of $\lambda = 3.04$, which yields the red curve shown in Figure~\ref{fig:numgiantimpacts}. This Poisson distribution function has a mean at 3.0 and a median of 3.4. While each of these giant impacts is likely to deplete a significant fraction of volatile content of the planet \citep[c.f.][]{Stewart.etal:2015}, most of these giant impacts are unlikely to fully strip an atmosphere and ocean. 

\subsection{Final giant impact time}
The \emph{final} giant impact received by an Earth-analog largely influences its dynamical state and bulk composition. With the relatively large number of simulations performed in this work, we can make probabilistic inferences from the distributions of the collision times of giant impacts onto Earth-analogs. 

\begin{figure*}
\includegraphics[width=0.9\textwidth]{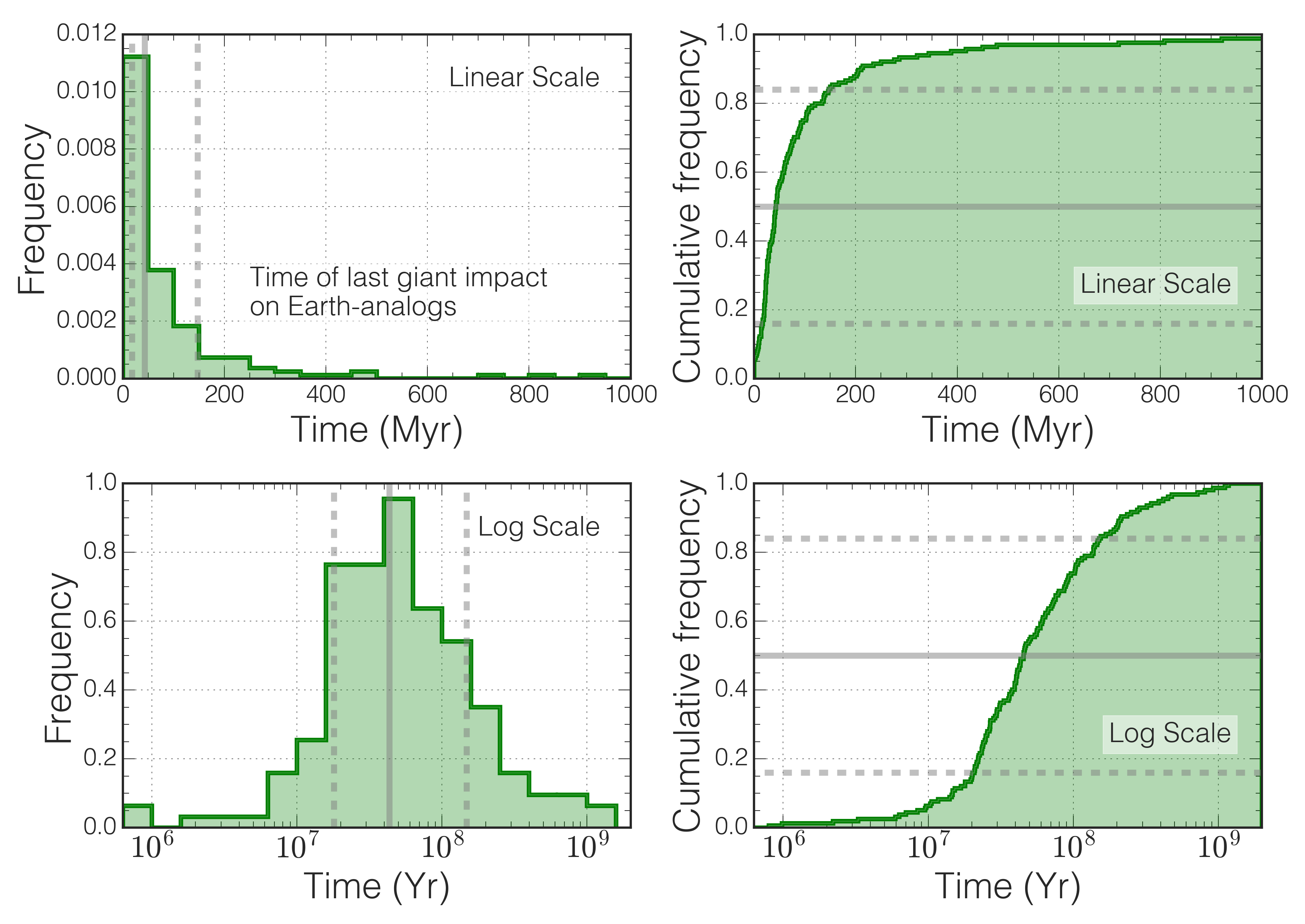}
\caption{
The distribution of times of last giant impact (defined by $Q_s > 2\times 10^6$ J kg$^{-1}$) for Earth-analogs that survive the duration of the simulation. The left panels show a histogram of the time of last giant impact with bin-width of 50 Myr. The right panels show the cumulative frequency of the same distribution. The gray dashed lines are drawn at the 15.9 and 84.1 percentiles and the gray solid lines are drawn at the median of the distribution. The 15.9, 50 and 84.1 percentiles are at 17 Myr, 43 Myr and 147 Myr after the beginning of the simulations, respectively. The upper panels show the data in linear space and the lower panels in log space.\label{fig:finalimp}
}
\end{figure*}

In Figure~\ref{fig:finalimp} we show the distribution of times of the final giant impact onto an Earth-analog in all of our fragmentation simulations. About 56\% of our Earth-analog planets experience their last giant impact within the first 50 Myr of the simulation and 75\% experience their final giant impact within the first 100 Myr. This mirrors our understanding of the range of feasible times for the Earth's Moon-forming impact (Figure~\ref{fig:timeline}). In terms of final giant impact times, we find that the Earth (assuming the Moon-forming impact was its final giant impact) is relatively typical.

The time of the final giant impact onto an Earth-analog is correlated with the abundance of late (post-impact) accreted mass by that Earth-analog, primarily because the amount of solid material available in the disk decreases with time as planets are growing. \citet{Jacobson:2014} illustrated this correlation by performing a large set of $N$-body perfect-accretion simulations using different models for the orbits of the giant planets. For simulations that included giant planets near their current orbits (comparable to our simulations), they found the fraction of mass accreted by an Earth-analog after its final giant impact ranged from about 10 -- 20\% for impacts occurring at 10 Myr to about 1\% for final impacts at around 100 Myr. 

From geochemical constraints, primarily the abundance of highly siderophile elements (HSEs) in the Earth's mantle compared to abundances found in chondritic meteorites, the fraction of mass that Earth accreted after the Moon-forming impact has been inferred to be less than 1\% \citep{Chou:1978, Walker:2009}. Assuming an estimate of the Earth's late accreted mass of 0.003 -- 0.006\mearth, \citet{Jacobson:2014} used their correlation to place constraints on the time of the Moon-forming impact to 95$\pm$32 Myr (for simulations that included giant planet migration) to a median time of about 110 Myr (for giant planets near their present orbits) after the start of the Solar System.  

In Figure~\ref{fig:latemass} we show, for all of our fragmentation simulations, a similar correlation between the final giant impact time for an Earth-analog and the fraction of mass the planet accreted after the impact. Compared to \citet{Jacobson:2014}, our results appear biased towards higher fractions of late mass accreted for a given final giant impact time. A direct comparison is difficult due to the differences in initial conditions and definitions. \citet{Jacobson:2014} began with half of the disk mass in about 100 bodies that are Moon-size to Mars-size and half of the disk in thousands of non-mutually interacting Ceres-sized (about 470 km) planetesimals. Our disk begins with half of the mass in 26 Mars-size embryos and the other half in 260 Moon-size planetesimals, and fragments are allowed during the simulation with a minimum resolution of about half a Moon mass. Additionally, our definition for a giant impact relies on a minimum specific impact energy that is physically motivated and more stringent, whereas \citet{Jacobson:2014} define a giant impact as a collision between an Earth-analog and a surviving Moon-size or Mars-size embryo (comparable to our initial planetesimals and embryos). Regardless of these differences, \citet{Jacobson:2014} assume that all of the late mass accreted has a chondritic composition, which would allow for a direct correlation between the mantle concentration of HSEs and the mass accreted after the Moon-forming impact. Our simulations show that some of the late accreted mass are fragments. Studies of earlier stages of planetesimal formation have shown that planetesimals are differentiated and exhibit a wide range of iron core fractions \citep{Carter.etal:2015}. Fragments created in the final stages of planet formation are therefore likely dominated by rocky mantle and should be depleted in HSEs compared to chondrites. As a result, more mass can be accreted post-giant impact than implied by the mantle's HSE abundance, so our results are consistent with the observations and uncertainty in the timing of the Moon-forming impact.

\begin{figure}
\plotone{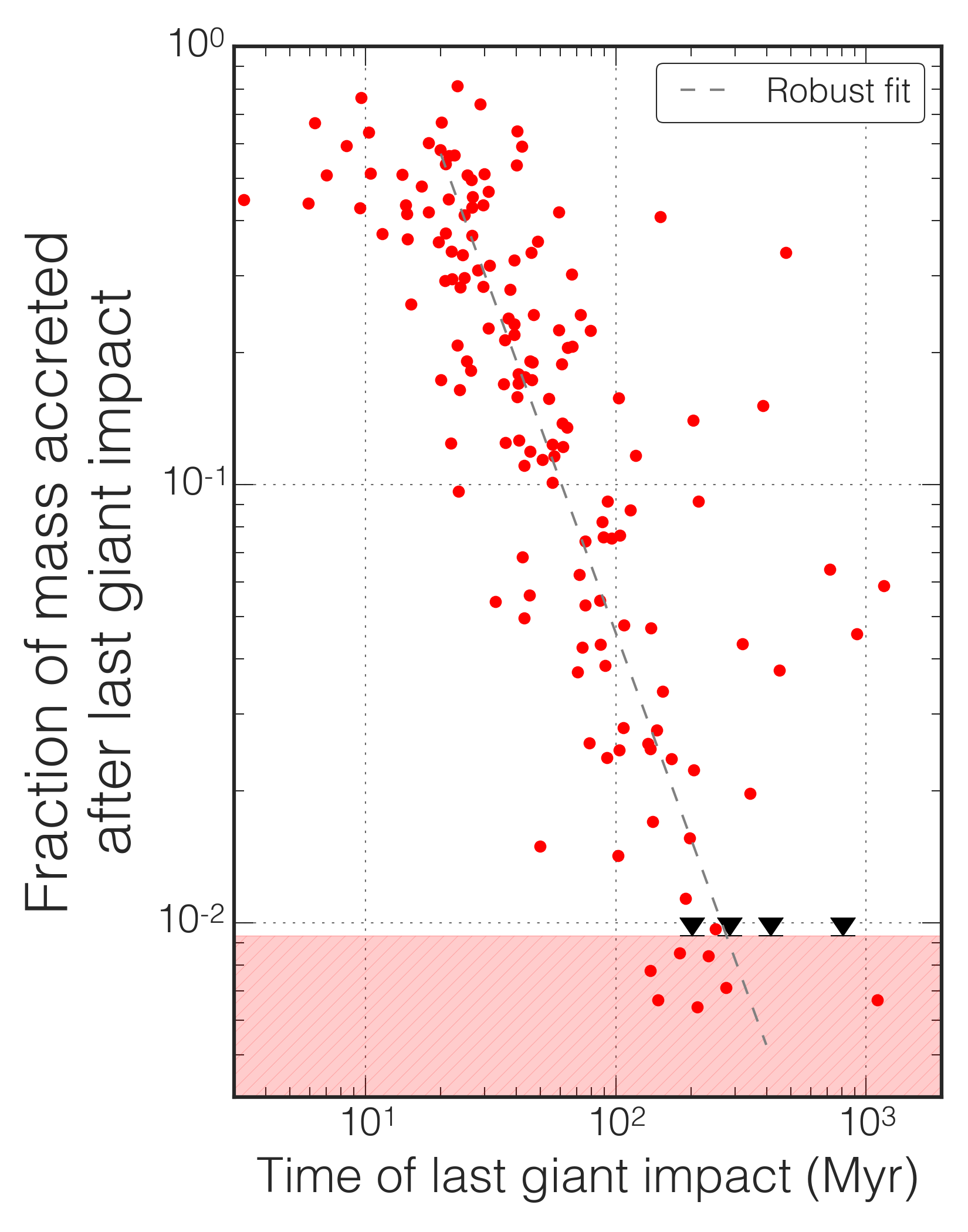}
\caption{
The proportion of material that was accreted by an Earth-analog after its last giant impact, plotted against the time of that last giant impact, is shown for all of the Earth-analogs that formed in our fragmentation simulations. More mass is accreted if the final impact occurs early \citep[following][]{Jacobson:2014}. The red hatched region shows where we have poor sensitivity because it is smaller than our initial minimum planetesimal mass of 0.009 \mearth. The black upper limit points show bodies where no additional material was accreted after the final giant impact. The gray line is a robust fit to the red data points for giant impacts after 20 Myr.\label{fig:latemass}
}
\label{fig:latemass}
\end{figure}

\begin{figure}
\plotone{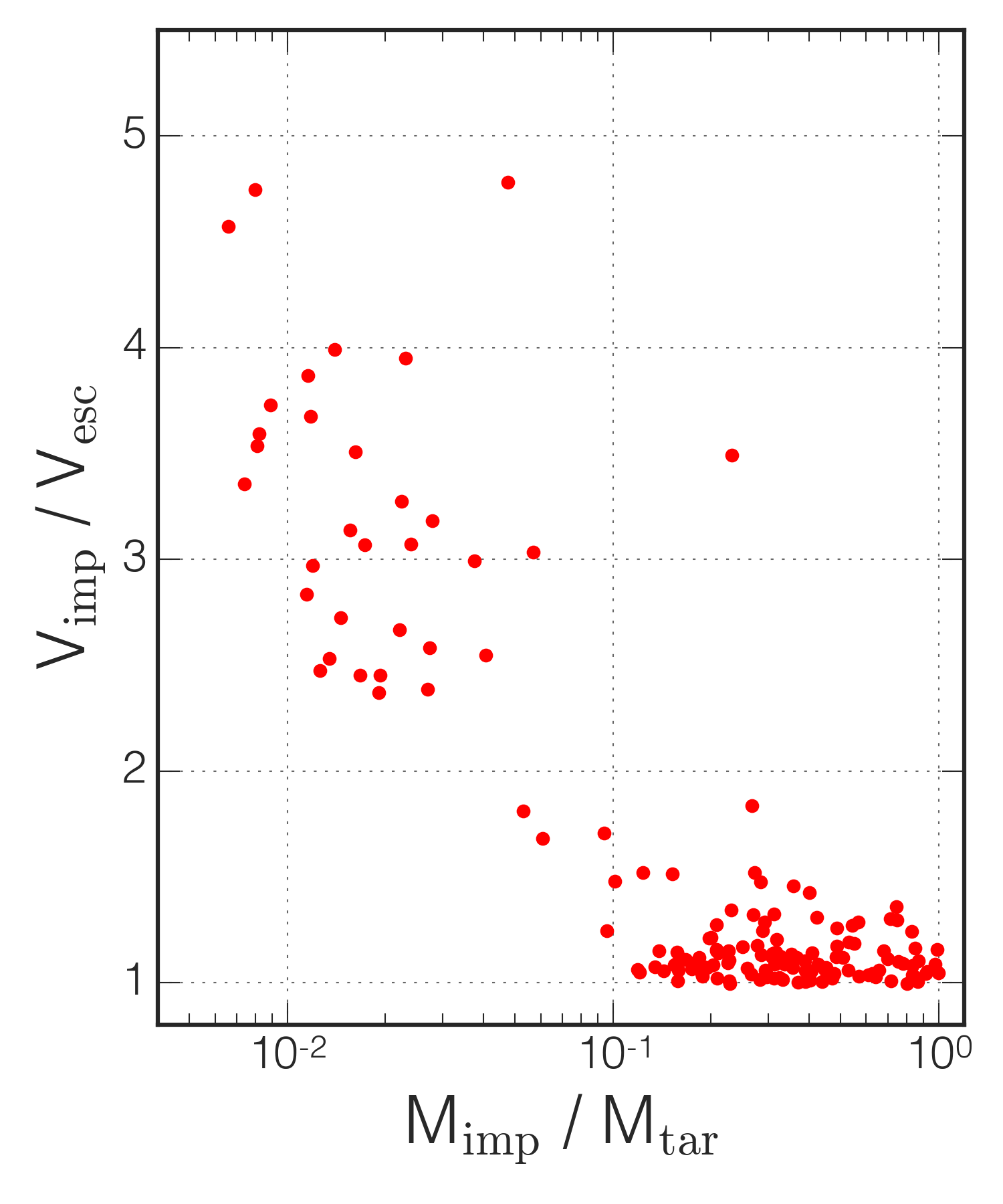}
\caption{For each final giant impact onto an Earth-analog, we show the impact velocity relative to the mutual escape velocity of the planets as a function of the mass ratio of the impacting body to the final planet. Smaller impactors onto more massive embryos have higher impact velocities compared to bodies of comparable mass.\label{fig:massratiovimp}}
\label{fig:massvimp}
\end{figure}

While our motivation in this work is not to model and constrain the timing of the Earth's Moon-forming impact, the distribution of mass ratios and impact velocities of the final giant impacts in our simulations  may shed some light on the likelihood of a given scenario and may be useful for comparing Moon-formation models. In Figure~\ref{fig:massvimp} we show the ratio of the impactor mass to the target mass as a function of the impact velocity (scaled by the mutual escape velocity) for all final giant impacts onto the Earth-analogs that formed in our fragmentation simulations. Most impact events that involve bodies of comparable mass have impact velocities that are 1 -- 2 times the mutual escape velocity, whereas smaller relative impactors reach impact velocities up to 5 times the mutual escape velocity. 

\subsection{On the validity of the initial conditions}\label{sec:limits} 
Here we briefly address some caveats that should be considered when interpreting these (and other similar $N$-body) results. Firstly, our initial conditions begin with objects that are roughly Moon-sized to Mars-sized. In reality the initial disk would contain these in addition to a significantly larger number of small bodies. In most $N$-body accretion simulations a much higher fidelity is usually desired but must be balanced with available computational resources.

The choice of the minimum fragmentation mass in our model constrains the total number of objects.  We selected a minimum fragmentation mass of a little under 0.5 a lunar-mass, the same value used in \citet{Chambers:2013}, so we could compare results and also perform a statistically large sample of simulations in a reasonable time. It is expected that the final planets that form, and their collision histories, will be different with the selection of a less massive (and more physically justified) minimum fragmentation mass. 
 
\begin{figure}
\plotone{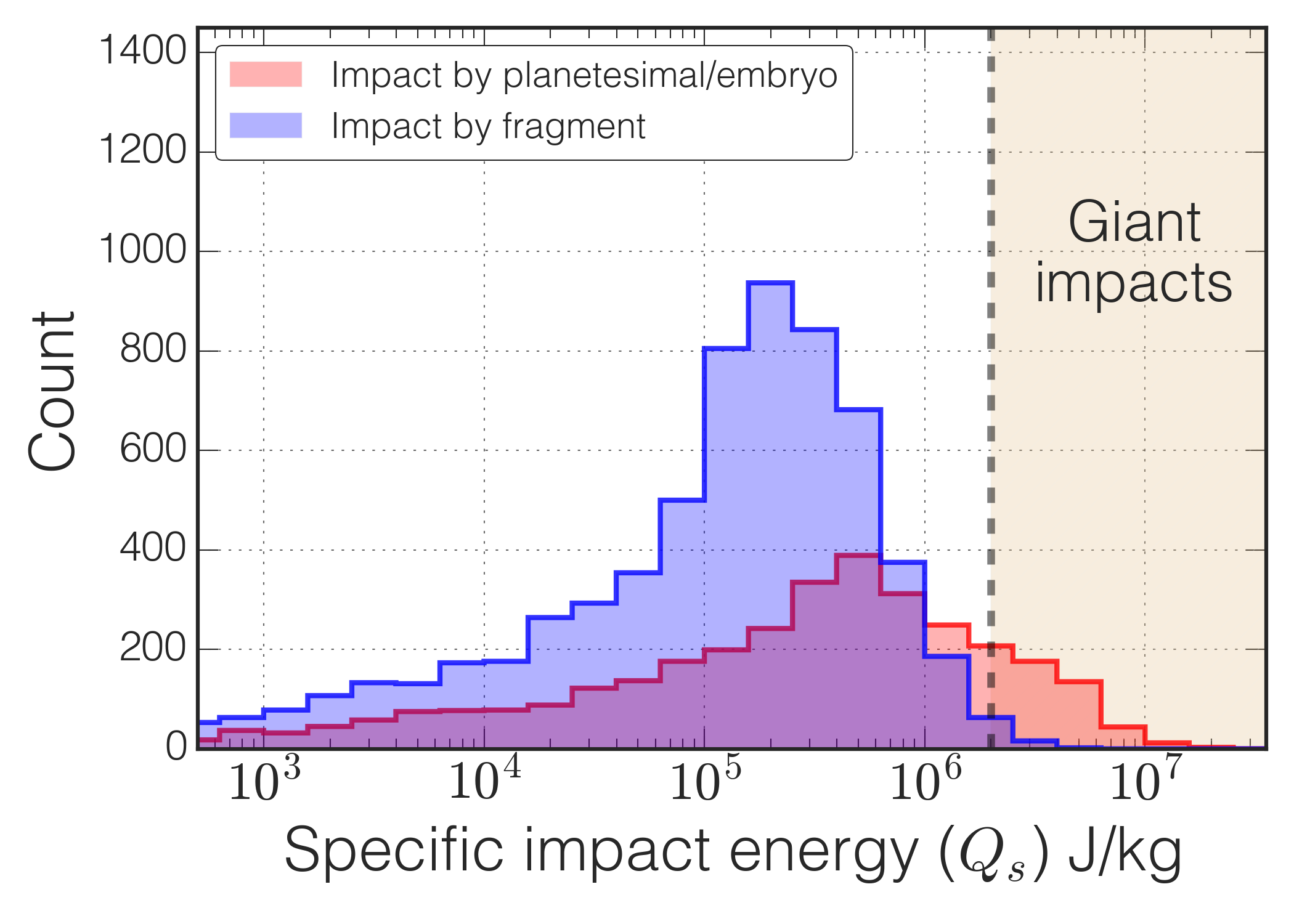}
\caption{Specific impact energies, $Q_s$, for all impactors onto the Earth-analogs that formed in our fragmentation simulations. The blue-shaded region represents material that had been previously fragmented, and the red-shaded region represents material that was at least as massive as our initial planetesimals (about the mass of the Moon). 93\% of the impactors that were involved in giant impacts (according to our definition as given in Figure~\ref{fig:Qslinear}) were by bodies that were non-fragments. Therefore, our minimum fragment size does not have a significant effect on our statistics of giant impacts.
\label{fig:impactor}}
\end{figure}

Of particular concern in this work was whether our conclusions on the numbers and times of giant impacts are biased by our minimum fragment mass. In Figure~\ref{fig:impactor} we show the specific impact energies of all impacts onto Earth-analogs split into two populations, (a) impactors more massive than one lunar mass and (b) impactors less massive than one lunar mass, where one lunar mass is approximately equal to the initial mass of the planetesimals in our simulations. We found that while only a third of total impactors have a mass of at least one lunar mass, 93\% of giant impacts are from bodies at least as large as our initial planetesimal mass. Therefore, our results are unlikely to depend strongly on the initial fragment mass since very few of the giant impacts are from fragments.

The initial conditions of the giant planets should be taken into consideration when interpreting the results presented herein. Classical models of the Solar System have assumed that the giant planets formed in situ \citep{Chambers:2001, Chambers:2013, Quintana.Lissauer:2014}. All of our simulations presented herein have adopted this model and include analogs to Jupiter and Saturn near their present orbits. Recent models have suggested that the giant planets may have originated on more eccentric orbits or may have undergone significant migration \citep{Raymond.etal:2009,Walsh:2012,Raymond.Morbidelli:2014}. Because these giant planets are the largest perturbers (not counting the central star), their configuration can significantly alter the growth of the terrestrial planets. For this article we focused on the classical theory of planet formation which includes Jupiter and Saturn at their present orbits. Exploring the effects of various giant planet configurations on the collision history of Earth-analogs would benefit Solar System models as well as planetary systems formed around other Sun-like stars in our galaxy.

Finally, we want to stress the importance of considering these processes in a statistical fashion and inferring results from probability distributions. Although planet formation in general is a chaotic process, accretion models have been fine-tuned for decades to reproduce and understand our Solar System. Instead, the current architecture of the planets in our Solar System should be thought of as one draw from the probability distribution of planetary systems that can be formed from the same protoplanetary disk. It is inference of this `generalized Solar System' from which our own planetary system is drawn that simulations such as those performed here are sampling. Caution should be used if drawing strong conclusions based on a small number of simulations.

\section{Summary and conclusions}\label{sec:conclusions}

We have presented a large set of numerical simulations of late stage terrestrial planet formation around a Sun-like star with outer planets analogous to Jupiter and Saturn perturbing the system. In total, 280 $N$-body simulations were performed, which is an order-of-magnitude larger number than nearly all previous numerical studies of this stage that used fully interacting bodies. Half of the simulations used a standard perfect-accretion model that treated all collisions as perfect mergers. The other half used an $N$-body algorithm that included a state-of-the-art collision model \citep{Leinhardt.Stewart:2012,Chambers:2013}, allowing a wider range of collision outcomes including fragmentation and hit-and-run impacts. All simulations began with virtually the same initial disk of hundreds of Moon to Mars-sized bodies.

We found that when fragmentation is enabled, the number of final planets and their mass distribution were consistent with those formed in the standard model, however the collision history of these planets differed significantly among the two sets. The typical accretion timescales were approximately doubled in the fragmentation simulations. The number of larger embryos and the total mass remaining as a function of time, however, were comparable in both models suggesting that much of the fragmented material was eventually re-accreted. In our fragmentation simulations, only about 64\% of all collisions lead to merging (and most of these mergers resulted in some fragmentation), validating the importance of including fragmentation when studying characteristics of terrestrial planet formation that depend on their collision history. 

We ran our fragmentation simulations to 2 Gyr to examine the collision history of the Earth-analogs, which we defined as planets that formed between 0.75 -- 1.5 AU from the Sun with a mass $>$ 0.5 \mearth. Of the 140 simulations that included fragmentation, 130 produced at least one Earth-analog and 34 produced two. To quantify the collisions that lead to the formation of an Earth-analog, we adopted a unit of specific impact energy $Q_s$ developed by \citet{Stewart.etal:2015, Lock.etal:2015} that takes into account the mass ratio and impact geometry for each collision. We then defined a threshold for a `giant impact' of $Q_s$ = $2\times 10^6$ J/kg, which is a value energetic enough to strip about half of a planet's atmosphere. This choice of $Q_s$ is also comparable to values of $Q_s$ estimated for recent (high-angular momentum) Moon-formation giant impact scenarios. All but seven of the Earth-analogs experienced at least one giant impact, according to our definition, and the average number of such impacts was 3.0 per planet within the 2 Gyr simulations. 

The $\emph{final}$ giant impact onto an Earth-like planet can shape its bulk composition and potential habitability. In our Solar System, the Earth's final giant impact was likely the Moon-forming giant impact \citep{Canup.Asphaug:2001, Canup:2004, Canup:2008,Canup:2012, Cuk.Stewart:2012}. Of the 96\% of Earth-analogs that experienced a giant impact, 20-30\% experienced their final giant impact after planet formation had essentially finished (100 -- 200 Myr into the simulation). The median final giant impact time for all Earth-analogs in our fragmentation simulations was 43 Myr with $\pm1\sigma$ equivalent intervals at 17 and 147 Myr, consistent with leading Moon-formation giant impact models developed for the Solar System. 

The timing of the emergence of life on Earth, and whether there were multiple origins due to life-sterilizing impacts, remains unclear despite the multitude of long-standing theories. The oldest known microfossils date the development of life to roughly 1 Gyr after the Solar System formed \citep{awramik} and there is indirect evidence for life occurring even earlier \citep{mojzsis96,mckeegan07,bella15}. We have shown with our 2 Gyr simulations that complete atmospheric removal from an Earth-analog planet in a single giant impact is rare. Just 15 of the 167 Earth-analogs in our simulations received an impact with an energy above 10$^7$ J/kg -- our full atmosphere stripping criterion, and no Earth-analog experienced an impact capable of stripping an entire ocean. Atmospheric loss is even more difficult if a planet lacks an ocean \citep{Genda.etal:2005}. Partial atmosphere stripping is much more common, however only 2 of our 167 Earth-analogs experienced such an impact after 1 Gyr. 

Given that the Earth's volatile content is relatively small compared to its mass \citep{Marty.Yokochi:2006}, events that remove a substantial fraction of an atmosphere may not necessarily be detrimental to habitability. If a planet has a thick atmosphere, these giant impacts may even help habitability by thinning out an atmosphere and avoiding runaway greenhouse effects. To explore the distribution of life-sterilizing impacts during late stage planet formation, higher resolution simulations than those presented here are needed. Impacts at energies an order of magnitude lower than the giant impact definition adopted here are capable of sterilizing an Earth-like planet \citep[e.g.][]{Canup:2008}. From our distributions of impact energies, it is clear that impacts that are potentially life-threatening are likely much more common than the bulk atmospheric removal events examined in this article.

\acknowledgments
The authors would like to thank Jack Lissauer, Billy Quarles, Chris Henze, Simon Lock and Sarah Stewart for comments that greatly improved this manuscript. E.V.Q is supported by a NASA Senior Fellowship at the Ames Research Center, administered by Oak Ridge Associated Universities through a contract with NASA. The simulations presented here were performed using the Pleiades Supercomputer provided by the NASA High-End Computing (HEC) Program through the NASA Advanced Supercomputing (NAS) Division at Ames Research Center.

\bibliographystyle{apj}
\bibliography{refstom}

\end{document}